# Bright hybrid excitons in molecularly tunable bilayer crystals

**Authors**:

Tomojit Chowdhury[1,2,10], Aurélie Champagne[3,4,10], Patrick Knüppel[5], Zehra Naqvi[6], Ariana Ray[7], Mengyu Gao[1], David A. Muller[7], Nathan Guisinger[8], Kin Fai Mak[5], Jeffrey B. Neaton[3,4,9], & Jiwoong Park[1,2,6]*

**Affiliations**:
[1]Department of Chemistry, University of Chicago, Chicago, IL, USA
[2]James Franck Institute, University of Chicago, Chicago, IL, USA
[3]Department of Physics, University of California, Berkeley, CA, USA
[4]Materials Science Division, Lawrence Berkeley National Laboratory, Berkeley, CA, USA
[5]Department of Physics, Cornell University, Ithaca, NY, USA
[6]Pritzker School of Molecular Engineering, University of Chicago, Chicago, IL, USA
[7]Department of Applied and Engineering Physics, Cornell University, Ithaca, NY, USA
[8]Center for Nanoscale Materials, Argonne National Laboratory, Lemont, IL, USA
[9]Kavli Energy Nanoscience Institute, University of California, Berkeley, CA, USA
[10]These authors contributed equally to this work

*corresponding author: jwpark@uchicago.edu

**Abstract**:

**Bilayer crystals, built by stacking crystalline monolayers, generate interlayer potentials that govern excitonic phenomena but are constrained by fixed covalent lattices and orientations. Replacing one layer with an atomically thin molecular crystal overcomes this limitation, as diverse functional groups enable tunable molecular lattices and interlayer potentials, tailoring a wide range of excitonic properties. Here, we report hybrid excitons in four-atom-thick hybrid bilayer crystals (HBCs), directly synthesized with single-crystalline perylene diimide (PDI) molecular crystal atop $WS_2$ monolayers. These excitons arise from a hybridized bilayer band structure, revealed by lattice-scale first-principles calculations, inheriting properties from both monolayers. They exhibit bright photoluminescence (PL) with near-unity polarization above and below the $WS_2$ bandgap, along with spectral signatures of exciton delocalization—supported by theory—while their energies and intensities are tuned by modifying the HBC composition by synthesis. Our work introduces a molecule-based 2D quantum materials platform for bottom-up design and control of optoelectronic properties.**

Conventional bilayer crystals are fabricated by stacking two mechanically exfoliated covalent two-dimensional (2D) monolayers with small differences in orientation or/and lattice constants, generating interlayer potentials[1]. The nanoscale tuning of periodicity of these potentials enables the realization of exotic phases of matter, including unconventional superconductors[2], insulators[3–5], and condensates[6,7]. However, in these bilayer systems, the interlayer potential is primarily dictated by the relative orientation of fixed covalent structures—predominantly honeycomb lattices—which imposes constraints on the tunability and symmetry of the resulting materials.

Hybrid bilayer crystals (HBCs), created by replacing a covalent monolayer with an atomically thin 2D molecular crystal, introduce additional tuning parameters that can overcome the structural and electronic constraints of conventional bilayers through direct synthesis[8]. A monolayer 2D molecular crystal, composed of discrete molecular units with well-defined dipoles, creates a periodic array of localized one-electron states[9–11], which can be fine-tuned through adjustments to lattice spacing, density, and symmetry—set by the molecular size, shape, and non-covalent intermolecular interactions, such as the extent of in-plane hydrogen bonding[12]. In addition, its localized electron wavefunctions, derived from the highest occupied and lowest unoccupied molecular orbitals (HOMOs and LUMOs) of individual monomers, form electronic bands with negligible dispersion and localized excitons[13,14]. Within an HBC, however, these molecular one-electron states may strongly interact with the delocalized states of the semiconducting covalent 2D crystal, such as monolayer transition metal dichalcogenides (TMDs), potentially leading to intra- and interlayer delocalization[15–18]. This interaction can produce a hybridized bilayer band structure with greater dispersion, enabling hybrid excitons that delocalize across the two layers while retaining molecule-like properties, such as strong optical polarization anisotropy[19], as schematically illustrated in **Fig. 1a**.

Previous studies have reported observations of interfacial excitons in molecule-based 2D heterostructures[18,20–23]. However, a bright excitonic PL from structurally precise bilayer crystals with hybrid character—where excitons inherit properties from both molecular and TMD monolayers—has yet to be realized. This limitation mainly stems from the absence of an atomically thin, single-crystalline bilayer materials platform, as well as the lack of a comprehensive lattice-scale first-principles framework to elucidate its band structure and excitonic behavior, both of which we address in this work.

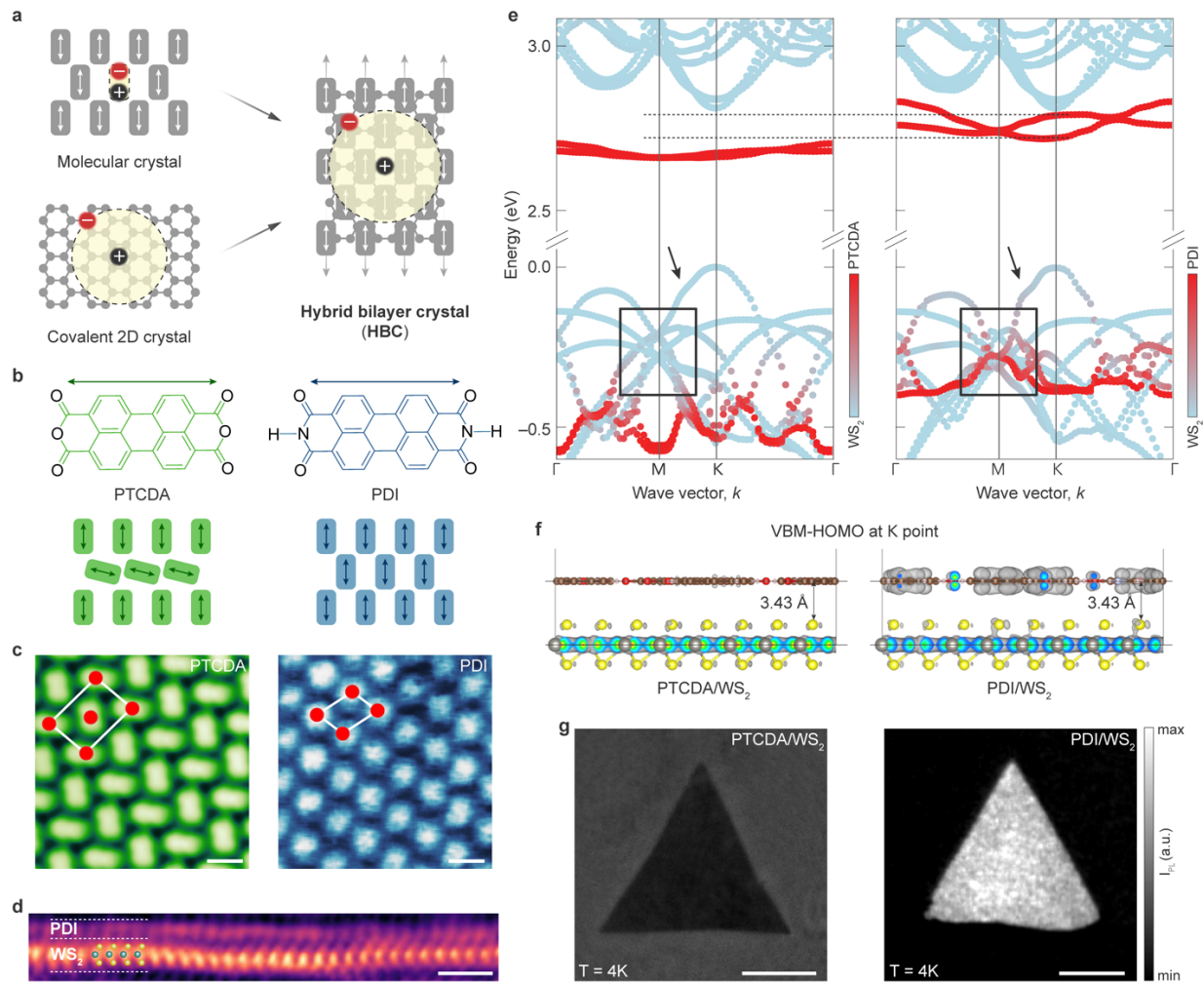

**Figure 1. HBC lattices, bilayer band structures, and bright PL emission.** (**a**) Schematic illustrating the formation of an HBC from individual molecular and covalent 2D crystals, along with their associated exciton spatial distributions. White (grey) double-sided arrows denote single-molecular (molecular crystal) polarization directions. (**b**) Chemical structures (top) of PTCDA and PDI molecules, with cartoon representations of their molecular orientation in 2D (bottom). Single-molecule dipole moments are indicated with double-sided arrows. (**c**) Room-temperature UHV STM images of PTCDA and PDI grown on monolayer $WS_2$. Molecular lattice points and unit cells are highlighted with red circles and white boxes, respectively. Scale bars: 1 nm. (**d**) Cross-sectional ac-STEM image of a $PDI/WS_2$ domain, with dashed white lines indicating the two monolayer boundaries. Colored circles mark W and S atoms. Scale bar: 5 Å. (**e**) GW-computed band structures of $PTCDA/WS_2$ (left) and $PDI/WS_2$ (right), highlighting orbital hybridization features by dashed lines, arrows, and boxes. Red (light blue) color denotes contributions from molecular ($WS_2$) electronic states. (**f**) DFT-calculated real-space squared wavefunctions of the VBM-HOMO (side-views) at K point for $PTCDA/WS_2$ (left) and $PDI/WS_2$ (right). Electron densities associated with W-$d$, C-$p$, and O-$p$ orbitals are highlighted in grey. (**g**) Wide-field polarized PL images of $PTCDA/WS_2$ (left) and $PDI/WS_2$ (right) single crystals acquired at a photon energy of 2.15 eV at T = 4K ($I_{PL}$: PL intensity). Scale bar: 20 μm.

**Molecularly tunable bilayer band structures and excitons in HBCs**

**Figure 1b** (top) illustrates the structures of two planar aromatic molecules we study in our experiments, PTCDA and PDI, which share a two-fold symmetric perylene core but differ only in their peri substitutions: –O– (anhydride in PTCDA, right) versus N–H (imide in PDI, left). These differences influence intermolecular interactions (e.g., in-plane hydrogen bonding), ultimately shaping their 2D molecular crystal geometries, as illustrated in **Fig. 1b** (bottom). To explore how structural differences affect excitonic behavior, we first synthesized single crystalline HBCs comprising monolayer molecular crystals of these building blocks grown atop monolayer $WS_2$ single crystals. The details of the wafer-scale synthesis and room-temperature optical characterization of different 2D molecular crystals have been discussed in our previous study[24]. The single-crystal HBCs studied here follow a similar growth protocol. Ultrahigh vacuum scanning tunneling microscopy (STM) reveals distinct long-range order, with PDI forming a brick-wall lattice and PTCDA adopting a characteristic herringbone arrangement (**Fig. 1c**). Cross-sectional aberration-corrected scanning transmission electron microscopy (ac-STEM) confirms an atomically sharp interface between PDI and $WS_2$ monolayers resulting in a total thickness of just four atomic layers (**Fig. 1d**).

**Figure 1e** compares the GW-computed bilayer band structures of $PTCDA/WS_2$ (left) and $PDI/WS_2$ (right) (see Methods for details of calculations)[25–27], revealing notable differences in the degree of electronic hybridization. The bands are color-coded by the relative orbital contributions: red (light blue) corresponds to molecular ($WS_2$) monolayers, while pink to hybridized states. The band structure of $PDI/WS_2$, in particular, cannot be predicted based on a simple superposition of those of its constituents, as it exhibits features that differ significantly from both the nearly flat bands of an isolated PDI monolayer (Extended Data Fig. 1a) and the characteristic dispersive bands of $WS_2$ monolayer (Extended Data Fig. 1b). First, the energies of bands with dominant HOMO and LUMO character (red) in the HBC undergo a massive renormalization relative to their counterparts in the molecular monolayer due to nonlocal dielectric screening from the $WS_2$ layer. Second, the HBC bands with dominant LUMO character show noticeable dispersion and a degeneracy lifting (dashed lines in **Fig. 1e**), evolving from a single state (in the isolated PDI monolayer) to a doublet (in the HBC lattice). Third, the HOMO bands with pronounced $WS_2$ character (pink; boxes at M point) and the $WS_2$ valence bands with additional shoulders (arrows close to K point) indicate strong orbital hybridization. In contrast, the $PTCDA/WS_2$ shows much flatter LUMO-derived bands and negligible change to the $WS_2$ valence bands at high-symmetry points. In PDI/WS$_2$, the greater extent of interlayer hybridization is also apparent in the valence band maximum (VBM)-HOMO wavefunctions at K point (**Fig. 1f**, right; grey shaded region), whereas it remains negligible in PTCDA/WS$_2$ (**Fig. 1f**; left). The band alignment predicted for PDI/WS$_2$ (**Fig. 1e**, see also **Fig. 4b**) indicates only a small energy offset between the PDI LUMO level and the WS$_2$ conduction band edge, which agrees well with that obtained from our transport measurements (Extended Data Fig. 2). This close band alignment suggests favorable conditions for interfacial electronic coupling[28], which further facilitates interlayer hybridization.

PDI/WS$_2$, with strongly hybridized band structures, display unexpected PL emission. **Figure 1g** compares the low-temperature (T = 4K) co-linearly polarized PL images of single-crystal PDI/WS$_2$ (bright PL) and PTCDA/WS$_2$ (dark), measured near the molecular emission maxima near 2.1 eV (see Methods and Extended Data Fig. 3 for experimental details). This bright PL from PDI/WS$_2$ above the WS$_2$ excitonic gap (~2 eV)[29] is counterintuitive, as PL from higher-energy excitons in PDI (or PTCDA) crystals is typically quenched or heavily suppressed due to the close spatial proximity of lower-energy WS$_2$ exciton states. Indeed, no such signal is observed from PTCDA/WS$_2$ (Extended Data Fig. 4a), consistent with the expected PL quenching[16]. A similarly bright above-gap PL is also observed from PDI/MoS$_2$ (Extended Data Fig. 5a).

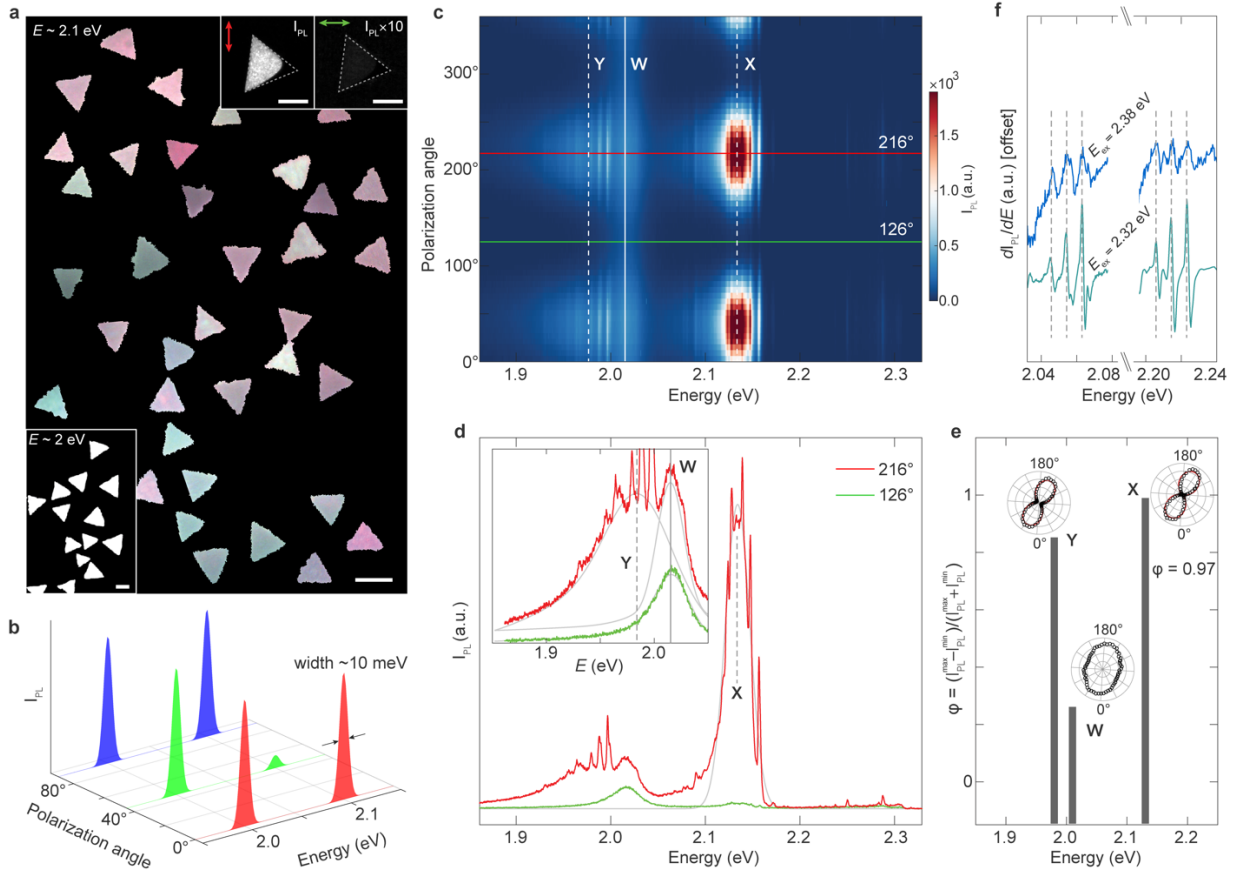

**Figure 2. Above-gap polarized PL in PDI/WS$_2$.** (**a**) Colinearly polarized PL RGB image of single-crystal islands ($E$ ~2.1 eV) at T = 4K. Scale bar: 20 μm. *Insets*: Co-polarized PL images of a partially grown PDI on WS$_2$ monolayer at two orthogonal polarization directions indicated as double-sided red and green arrows (top right); PL RGB image of single-crystal islands ($E$ ~ 2 eV) (bottom left). Scale bar: 10 μm. (**b**) A 3D plot illustrating polarization angle-dependent PL RGB imaging scheme used in (**a**) to distinguish anisotropic ($E$ ~2.1 eV) and isotropic ($E$ ~2 eV) single crystal domains with ~10 meV spectral resolution (polarization angle = angle of the half-wave plate). (**c**) Angle-dependent confocal PL map from a single-crystal island highlighting three peaks—X, Y (dotted white lines), and W (solid white line) with ~1 meV spectral resolution. (**d**) PL line spectra generated from the linecuts in (**c**) at two orthogonal polarization angles. Raw data shown with Gaussian overlays (grey line) highlighting X with a dashed line. *Inset*: Zoomed-in spectral range at and below the WS$_2$ gap, highlighting peaks W and Y with solid and dashed lines, respectively, while Gaussian overlays are shown as grey lines. (**e**) PL anisotropy (φ) for peaks X, Y, and W is plotted as grey bars as a function of energy, with corresponding polar scatter plots for X, Y, and W derived from (**c**), overlaid with cosine-squared fits (red lines) for X and Y. (**f**) $dI_{PL}/dE$ vs. energy plot for two excitation energies ($E_{ex}$) 2.38 eV (blue) and 2.32 eV (teal), highlighting the sharp spectral features with dashed grey lines.

**Above-gap excitonic PL in PDI/WS$_2$ with near-unity PL polarization**

The bright above-gap PL signal exhibits a full linear polarization as shown in **Fig. 2a** (top inset), which compares PL images taken with parallel and perpendicular polarizations. **Figure 2a** (main panel) further shows an RGB false-colored wide-field PL image of PDI/WS$_2$ single crystals, revealing a distinct and spatially uniform PL anisotropy direction for each single-crystal domain (see also Supplementary Video 1). The RGB image in **Fig. 2a** is constructed by recording colinearly polarized PL intensity (I$_{PL}$) images at $E$ ~2.1 eV for three polarization orientations (0º, 45º and 90º), and then mapping them onto red (R), green (G), blue (B) channels, as schematically illustrated in **Fig. 2b** (see also Extended Data Fig. 3b). In contrast to these colorful triangles, applying the same imaging scheme to the WS$_2$ excitonic PL ($E$ ~2 eV) produces white triangles (bottom inset; **Fig. 2a**), demonstrating significantly more isotropic emission at this energy.

The polarization angle-dependent PL spectral map (**Fig. 2c**) and line spectra (**Fig. 2d**) acquired from a single-crystal PDI/WS$_2$ highlight three major PL peaks, denoted as X, Y, and W. Among these, peak X ($E_X$ = 2.13 eV; bandwidth = 34 meV), which corresponds to the PL signal in **Fig. 1g** (right) and **Fig. 2a**, is the brightest, with its energy close to the molecular (PDI) absorption energy (Extended Data Fig. 3c). In contrast, peak W ($E_W$ = 2.01 eV; bandwidth = 50 meV) shows significantly lower PL intensity, attributed to the neutral excitonic PL in monolayer WS$_2$ (inset; **Fig. 2d**)[29]. Additionally, a broad peak, Y ($E_Y$ = 1.98 eV; bandwidth = 55 meV), appears just below W (inset; **Fig. 2d**). Peaks X and Y both displays strong PL polarization anisotropy, φ, where φ = $((I_{PL}^{max} - I_{PL}^{min})/(I_{PL}^{max} + I_{PL}^{min}))$, with $I_{PL}^{max}$ and $I_{PL}^{min}$ representing the maximum and minimum PL intensities, respectively. Both X (φ = 0.97) and Y (φ = 0.83) exhibit strong anisotropic response with X showing near-unity anisotropy (**Fig. 2e**).

In addition, we observe two groups of sharp Raman peaks (2–3 meV bandwidths) around X and Y (but not W) in **Fig. 2c** and **Fig. 2d**, indicative molecular signatures. These Raman modes are characteristic of the aromatic C=C vibrations within the perylene core of PDI, in line with previous works[30,31]. This assignment was confirmed by the PL measurements at two excitation energies, 2.32 eV and 2.38 eV, which show that the sharp peaks shift by the ~63 meV excitation energy difference while the PL energies remain the same (**Fig. 2f**, Extended Data Fig. 6).

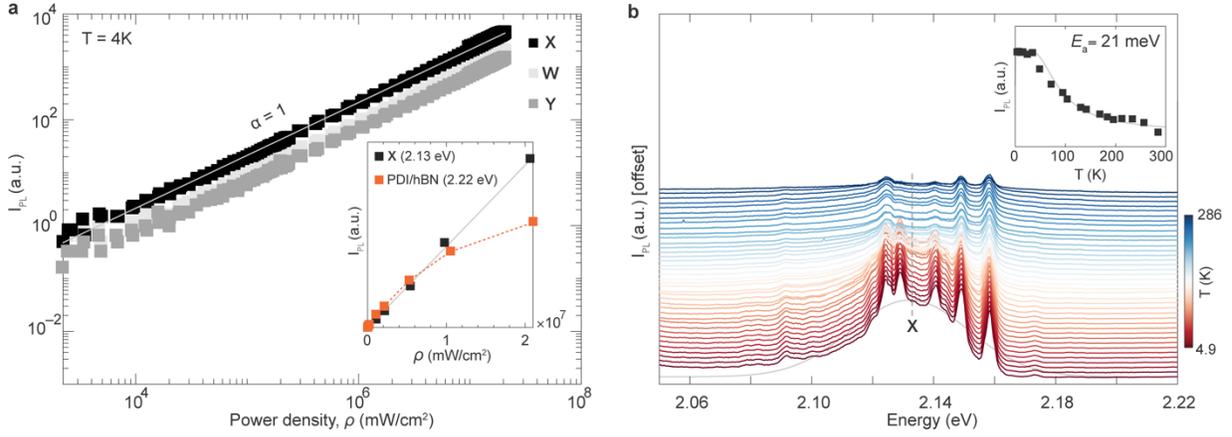

**Figure 3. Power- and temperature-dependent PL in PDI/WS$_2$.** (**a**) Logarithmic plot of excitation power-density ($\rho$) dependent I$_{PL}$ for X (black), Y (dark grey), and W (light grey) at T = 4K, with an overlaid solid grey line indicating a power-law fit of the form I$_{PL} \propto \rho^\alpha$ with $\alpha$ = 1. *Inset*: Comparison of $\rho$-dependent I$_{PL}$ for peak X (black) and that from a PDI/hBN flake (orange; $E$ = 2.22 eV). The solid grey line represents a linear fit (slope = 1), and the orange dotted line serves as a visual guide. (**b**) Temperature-dependent PL spectra highlighting peak X with a dashed grey line (Gaussian overlay shown in grey). *Inset*: Plot of I$_{PL}$ vs. temperature for peak X fitted with a single-exponential function (grey line). $E_a$, the thermal activation energy for X, is estimated from the exponential fit.

**Resemblance of anisotropic molecular PL signals to WS$_2$ excitons**

The two peaks, X and Y, exhibit characteristics of both molecular and TMD layers, suggesting that they originate from hybrid excitons. They show molecular signatures, including the near-unity linear polarization anisotropy and distinct aromatic Raman features. Additionally, their spectral properties—including peak widths, power and temperature dependencies—resemble those of TMDs while remaining distinct from standalone molecular crystals, as discussed below.

First, their PL bandwidths (~35–55 meV) are several times broader than that of PDI alone (~10 meV), as observed in PDI monolayers synthesized directly on hBN ($E$ = 2.22 eV; Extended Data Fig. 7) but remain close to that of W (**Fig. 2d**). Second, their excitation power-density ($\rho$) dependence differs markedly from the main signal from PDI/hBN, which saturates at high power (**Fig. 3a**, inset), indicating molecular (localized) excitonic behavior[32], whereas both X and Y in PDI/WS$_2$ remain linear (**Fig. 3a**), mirroring the behavior of W[33]. Finally, temperature-dependent PL measurements (**Fig. 3b** and Extended Data Fig. 4b) yield a thermal activation energy ($E_a$) of 21 meV for X in PDI/WS$_2$ extracted by using a single Arrhenius-like model (**Fig. 3b**, inset; see Methods). It is approximately an order of magnitude smaller than in freestanding perylene derivatives[34]; instead, it is similar to the binding energy of trions in monolayer TMD[35]. Together, these findings highlight a close resemblance of X and Y to W.

Moreover, we observe bright PL peaks associated with hybrid excitons in PDI/MoS$_2$, as shown in Extended Data Fig. 5. We observe an above-gap, polarized peak X ($E_X$ = 2.12 eV; $\varphi$ = 0.78) and a near-gap peak Y that appears above ($E_Y$ = 1.96 eV; $\varphi$ = 0.71), rather than below, the isotropic MoS$_2$ excitonic PL peak (denoted as M with $E_M$ = 1.91 eV; $\varphi$ = 0.23)[36]. This shows that the intensity, energy, and anisotropy of X and Y—observed only in PDI-containing HBCs—are controlled by the composition of either or both molecular and TMD monolayers, as expected for hybrid excitons.

The observation of bright hybrid excitons only in PDI-containing HBCs is attributed to the strong hybridization and orbital overlap between the HOMO and VBM, as shown in **Fig. 1e**. An experimental confirmation would require correlated PL and band structure measurements. While such measurements are outside of the scope of this paper, a recent photoemission study of PTCDA/WSe$_2$ reported below-gap hybrid excitons formed via resonant energy transfer[20].

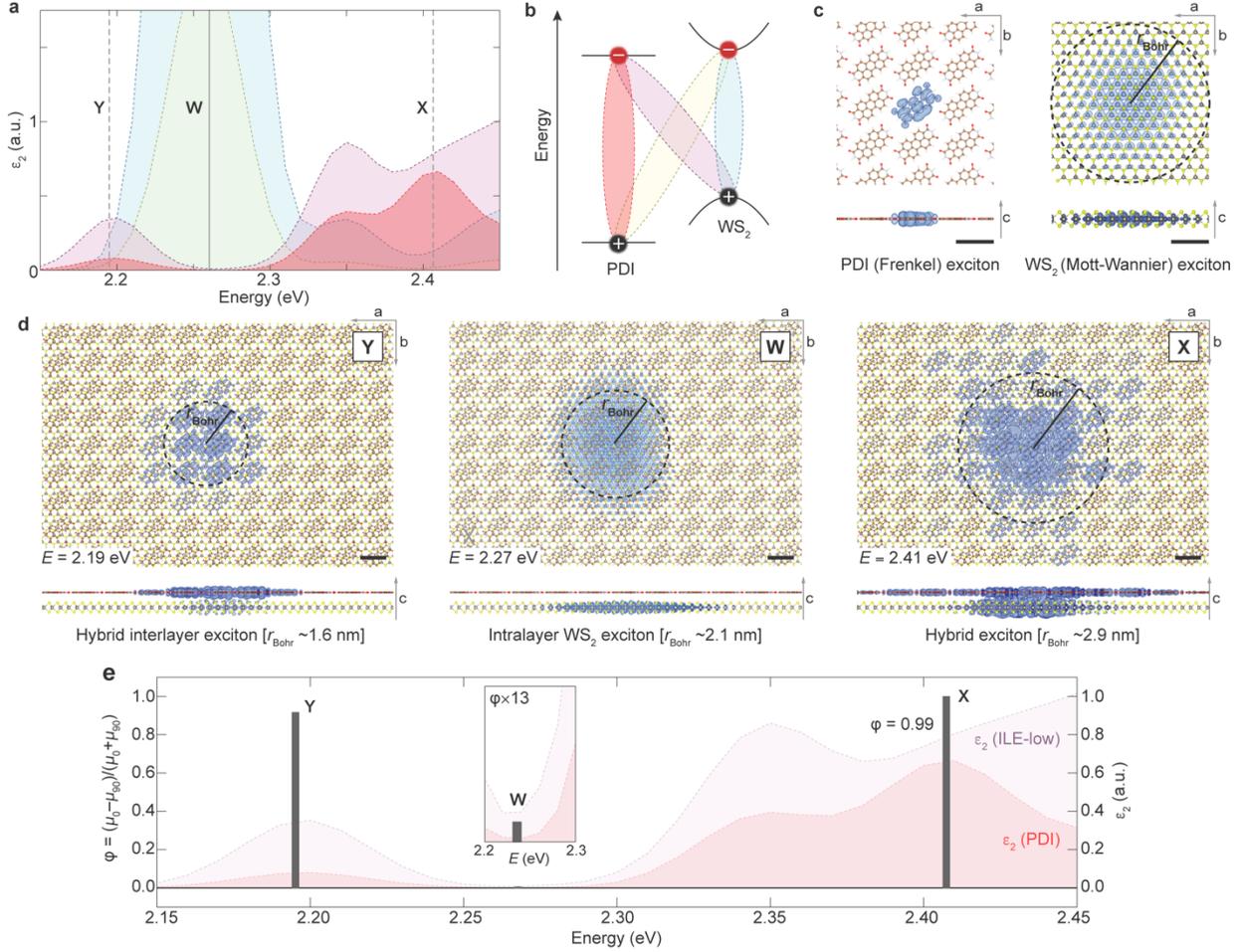

**Figure 4. Hybrid excitons in PDI/WS$_2$.** (**a**) Computed imaginary components of the dielectric function ($\varepsilon_2$) of the HBC, plotted against energy, showing contributions from excitons predominantly associated with PDI (red), WS$_2$ (cyan), and interlayer excitations (ILE) at low (pink) and high (yellow) energies. Peaks X and Y are marked with dashed lines, and W with a solid line. (**b**) PDI/WS$_2$ band alignment derived from the GW-band structure calculation, highlighting four excitonic species with different intralayer (red, cyan) and interlayer (pink, yellow) characters. (**c**) GW-BSE isosurface maps of the exciton wavefunctions (selected to include 98% of the electron density), showing the electron probability density (blue) for the most probable hole positions. Projections are shown on the *a-b* plane and along the *c*-axis for a Frenkel exciton (left) and a Mott-Wannier exciton (right) in freestanding PDI and WS$_2$ monolayers ($r_{Bohr}$ = Bohr radius). Scale bar: 1 nm. (**d**) GW-BSE projections onto a 20×20×1 supercell consisting of PDI and WS$_2$ monolayers together, highlighting excitons X (right), W (middle), and Y (left), each labeled with their respective energies and $r_{Bohr}$. Scale bar: 1 nm. (**e**) Anisotropy in oscillator strength ($\varphi$), computed from the exciton oscillator strengths for parallel ($\mu_0$) and perpendicular ($\mu_{90}$) polarizations, plotted as grey bars against energy. Peaks X, Y, and W are highlighted, with overlaid $\varepsilon_2$ traces for PDI (red) and low-energy ILE (pink) contributions.

**First principles insights into hybrid excitons in PDI/WS$_2$**

Our *ab initio* GW and Bethe-Salpeter equation (BSE) calculations fully support the origin of X and Y as hybrid excitons. **Figure 4** specifically shows the results for PDI/WS$_2$ (see Methods and Extended Data Fig. 1c for details). **Figure 4a** plots four components of the imaginary part of the dielectric function, $\varepsilon_2$, directly related to the linear absorption spectrum for PDI/WS$_2$, each component corresponding to transitions associated with PDI (red), WS$_2$ (cyan), and interlayer excitations (ILE) at low (pink) and high (yellow) energy states (see schematic in **Fig. 4b**)[37]. We identify three main features from our calculations at 2.19 eV, 2.27 eV, and 2.41 eV with their energy ordering consistent with the measured PL peaks: X appearing above W, and Y slightly below W (**Fig. 2c**).

To further analyze the spatial characteristics of the hybrid excitonic states, **Fig. 4c** first shows the real-space electron-hole correlation function of the primary excitonic states for freestanding PDI and WS$_2$ monolayers[38,39]. The former shows typical Frenkel-like (localized) excitons, with the latter showing Mott-Wannier-like (delocalized) excitons. **Figure 4d** plots similar GW-BSE maps for three distinct excitonic states in PDI/WS$_2$, corresponding to the spectral features X, Y, and W. Excitons X and Y both reveal significant in-plane delocalization over several unit cells within the HBC, with estimated exciton Bohr radii ($r_{Bohr}$) of 2.9 nm for X and 1.6 nm for Y, both comparable to that of W ($r_{Bohr}$ = 2.1 nm). In particular, the map for W in **Fig. 4d** is nearly identical to that in **Fig. 4c** (right), both showing a Mott-Wannier-like intralayer exciton confined entirely within the WS$_2$ monolayer.

In contrast, the vertical distributions of the exciton wavefunctions differ markedly for X and Y. Their electron densities (calculated with a hole fixed in WS$_2$) are vertically delocalized across both the WS$_2$ and PDI layers, suggesting that they are hybrid excitons. Combining these maps with our analysis of different transitions in **Fig. 4a** reveals the subtle differences in X and Y. For X, the wavefunction is significant in both layers and it is associated with both intra- and interlayer transitions (red and pink in **Fig. 4a**). For Y, however, the electron density is mostly confined within the PDI layer, separated from the fixed hole in WS$_2$, and it is primarily associated with an interlayer transition (pink in **Fig. 4a**). This makes Y similar to charge-transfer excitons observed in other related systems[21,40].

Moreover, our GW-BSE analysis predicts a strong polarization anisotropy for X and Y, consistent with our PL experiments. **Figure 4e** plots the linear polarization anisotropy, $\varphi = ((\mu_0 - \mu_{90})/(\mu_0 + \mu_{90}))$, calculated for each exciton based on the oscillator strengths for parallel ($\mu_0$) and perpendicular ($\mu_{90}$) polarizations. Exciton X, which has the highest contribution from PDI (red, **Fig. 4a**), displays near-unity polarization anisotropy ($\varphi = 0.99$), whereas W shows negligible polarization dependence. Y, with the largest ILE contribution (pink, **Fig. 4a**), also exhibits significant anisotropy ($\varphi \sim 0.9$). These computed values are comparable to the PL anisotropy observed experimentally (**Fig. 2e**). Altogether, our experiments and theory suggest that the polarization anisotropy is highly sensitive to the molecular contribution to the overall excitonic response.

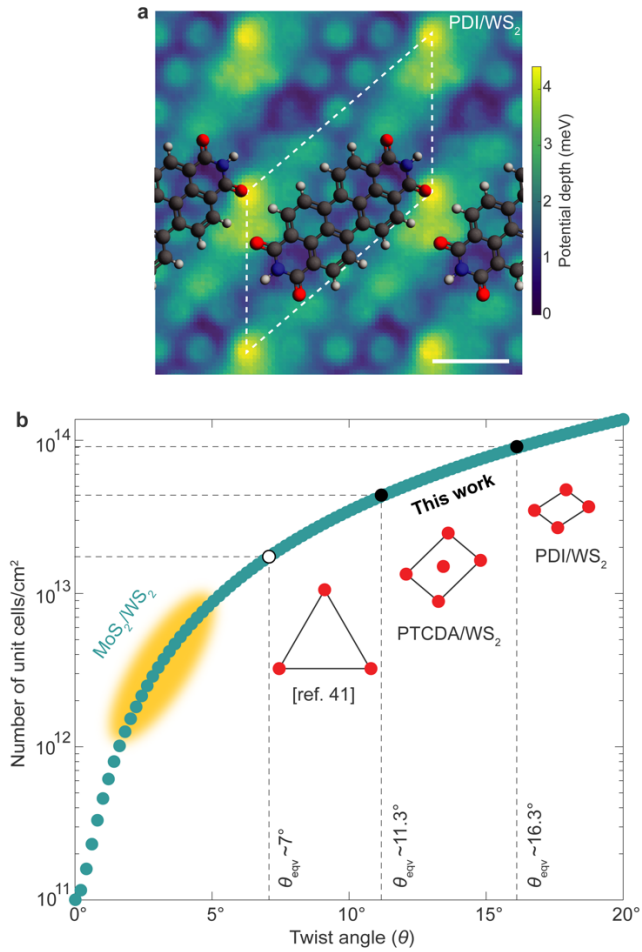

**Figure 5. HBCs for imprinting potential energy landscapes.** (**a**) Computed interlayer potential depth induced by the undoped PDI monolayer on the $WS_2$ surface, with overlaid PDI molecular structures. Dashed white lines highlight the periodicity of the potential energy landscape. Scale bar: 1 nm. (**b**) Numerically calculated lattice density, expressed as the number of unit cells/cm$^2$, plotted against interlayer twist angle ($\theta$) for a bilayer $MoS_2/WS_2$ superlattice. Dashed lines mark equivalent twist angles ($\theta_{eqv}$) and projected lattice densities for PDI/$WS_2$ and PTCDA/$WS_2$, with their corresponding $\theta_{eqv}$ values shown alongside. Solid circles represent our data from this work, while open circle indicates a molecular building block from literature[41], respectively. The low-density regime is highlighted in orange.

**Conclusions**

The above-gap hybrid exciton observed at low temperatures, particularly in PDI-based HBCs, highlights the need to explore the complex interplay between the molecular lattice and the TMD, incorporating contributions from phonons[42]. Indeed, HBCs introduce an interlayer potential energy landscape onto TMD shaped by the molecular monolayer, as demonstrated by our *ab initio* calculations. **Figure 5a** shows that even an undoped PDI monolayer can induce potential wells with an average depth of 4.5 meV at the $WS_2$ surface. Although shallower than in all-inorganic moiré bilayers[43], such a depth is still predicted to create a distinct charge localization pattern aligned with the molecular lattice and exciton wavefunction (white dashed line in **Fig. 5a**; see Extended Data Fig. 8). This effect could be further amplified through charge modulation and doping.

As the molecular crystals can be synthetically controlled to form low or high density lattices, our HBCs could provide an additional knob for engineering periodic potentials, in addition to an interlayer twist in covalent bilayers[44]. To demonstrate this concept, we plot the lattice densities of our HBCs and compare them with those of a $MoS_2/WS_2$ bilayer with various twist angles ($\theta$) in **Fig. 5b** (see Methods for calculation). While HBCs shown here have relatively large equivalent twist angles ($\theta_{eqv}$), extending our synthesis to other molecular motifs[41] could yield larger unit cells, enabling smaller $\theta_{eqv}$ and, in turn, access to a critical low-density regime[45] (orange shaded region in **Fig. 5b**). This approach unlocks new possibilities for molecularly tunable 2D quantum materials, where precise control over interlayer potentials and excitonic interactions could drive advances in next-generation quantum optoelectronic devices.

**Acknowledgements:** We thank Ce Liang for assisting with MOCVD. We also thank Zhen-Fei Liu, Olugbenga Adeniran, and Sahar Sharifzadeh for helpful discussions and for providing the scripts used for the calculations.

**Funding:** Primary funding for this work comes from the Air Force Office of Scientific Research (FA9550-21-1-0323) and the MURI project (FA9550-18-1-0480). Additional funding was provided by the Office of Naval Research (N000142212841) and Samsung Advanced Institute of Technology. T.C. acknowledges the Kadanoff-Rice Postdoctoral Fellowship from the National Science Foundation (NSF) Materials Research Science and Engineering Center under grant no. DMR-2011854. F.M. acknowledges support by the NSF Graduate Research Fellowship Program under grant no. DGE-1746045. Work performed at the Center for Nanoscale Materials, a U.S. Department of Energy (DOE) Office of Science User Facility, was supported by the U.S. DOE, Office of Basic Energy Sciences, under contract no. DE-AC02-06CH11357. The calculation in this work are primarily supported by the Center for Computational Study of Excited-State Phenomena in Energy Materials, funded by the U.S. DOE under contract no. DE-FG02-07ER46405. The Theory of Materials FWP at LBNL, funded by the DOE under contract no. DE-AC02-05CH11231, supported the development of the theories and models. Computational resources are provided by the National Energy Research Scientific Computing Center. P.K. acknowledges financial support from the Swiss National Science Foundation.

**Author Contributions:** T.C. and J.P. conceived the main idea of the project. A.C. performed the *ab initio* calculations and analysis of the excited states under the supervision of J.B.N. T.C. synthesized the materials with help from Z.N. T.C. and P.K. conducted optical measurements with help from Z.N. A.R. performed ac-STEM. N.G. assisted with STM. M.G. conducted additional transport measurements. T.C. analyzed all experimental data, developed numerical calculation and image analysis codes with help from Z.N. T.C. and J.P. wrote the paper with input from co-authors.

**Competing interests:** The authors declare no competing financial interests.

**Data and materials availability:** The code used to fold/map/add the noninteracting polarizability matrices was first proposed by Liu *et al.*[46] and is available upon request. The code used to decompose the excited states was first proposed by Ud Din *et al.*[37] and is available upon request. The code used to compute the electron-hole correlation function was first proposed in Sharifzadeh *et al.*[38] and is available upon request. The BerkeleyGW package used for GW-BSE calculations is a free, open source package available at berkeleygw.org. The code used to plot the band projections is available at pymtagen.electronic_structure package.